\documentclass[twocolumn,aps,prl,epsf,showpacs,superscriptaddress]{revtex4}
\usepackage{amsmath}
\usepackage{amsfonts}
\usepackage{epsfig}
\usepackage{epsf}
\usepackage{array}

\begin{document}

\title{Correlation effects in (111) bilayers of perovskite transition-metal oxides}
%\title{Correlation-induced phase transitions in (111) bilayers of perovskite transition-metal oxides}

\author{Satoshi Okamoto}
\altaffiliation{okapon@ornl.gov}
\affiliation{Materials Science and Technology Division, Oak Ridge National Laboratory, Oak Ridge, Tennessee 37831, USA}
\author{Wenguang Zhu}
\affiliation{ICQD/HFNL, University of Science and Technology of China, Hefei, Anhui 230026, China}
\author{Yusuke Nomura}
\author{Ryotaro Arita}
\affiliation{Department of Applied Physics, the University of Tokyo, Hongo, Bunkyo-ku, Tokyo 113-8656, Japan}
\author{Di Xiao}
\affiliation{Department of Physics, Carnegie Mellon University, Pittsburgh, Pennsylvania 15213, USA}
\author{Naoto Nagaosa}
\affiliation{Department of Applied Physics, the University of Tokyo, Hongo, Bunkyo-ku, Tokyo 113-8656, Japan}
\affiliation{RIKEN Center for Emergent Matter Science (CEMS), Wako, Saitama 351-0198, Japan}

\begin{abstract}
We investigate the correlation-induced Mott, magnetic and topological phase transitions 
in artificial (111) bilayers of perovskite transition-metal oxides LaAuO$_3$ and SrIrO$_3$ 
for which the previous density-functional theory calculations predicted topological insulating states. 
Using the dynamical-mean-field theory with realistic band structures and Coulomb interactions, 
LaAuO$_3$ bilayer is shown to be far away from a Mott insulating regime, and a topological-insulating state is robust. 
On the other hand, SrIrO$_3$ bilayer is on the verge of an orbital-selective topological Mott transition and 
turns to a trivial insulator by an antiferromagnetic ordering. 
Oxide bilayers thus provide a novel class of topological materials for which the interplay between 
the spin-orbit coupling and electron-electron interactions is a fundamental ingredient. 
\end{abstract}


\maketitle

\date{\today }

%\draft

\section{I. Introduction}

Topological insulators (TIs) are novel quantum states of matter 
characterized by the nontrivial band topology due to the relativistic spin-orbit coupling (SOC) \cite{Kane05,Bernevig06,Moore07,Fu07,Konig07,Hsieh08,Xia09}. 
While the TI states are essentially single-particle phenomena, 
the interplay between the SOC and strong Coulomb interactions in electronic systems has gained considerable attention 
\cite{Shitade09,Pesin10,Chaloupka10}. 
A variety of effects between these two couplings has been theoretically explored, 
such as TI states induced by Coulomb interactions \cite{Raghu08,Zhang09}, 
competition and/or cooperation between the two in 
a Kane-Mele-Hubbard (KMH) model\cite{Rachel10,Zheng11,Wen11} 
or 
a Bernevig-Hughes-Zhang (BHZ) model \cite{Yoshida13,Miyakoshi13,Budich13}, 
and topological Kondo insulators \cite{Dzero10,Wener13,Deng13}. 
%
%However, until now physical realization for such correlation-related phenomena has been limited to 
%an $f$ electron system SmB$_6$.\cite{Wolgast13,Kim13,XZhang13} 

\begin{figure}[tbp]
\includegraphics[width=0.9\columnwidth,clip]{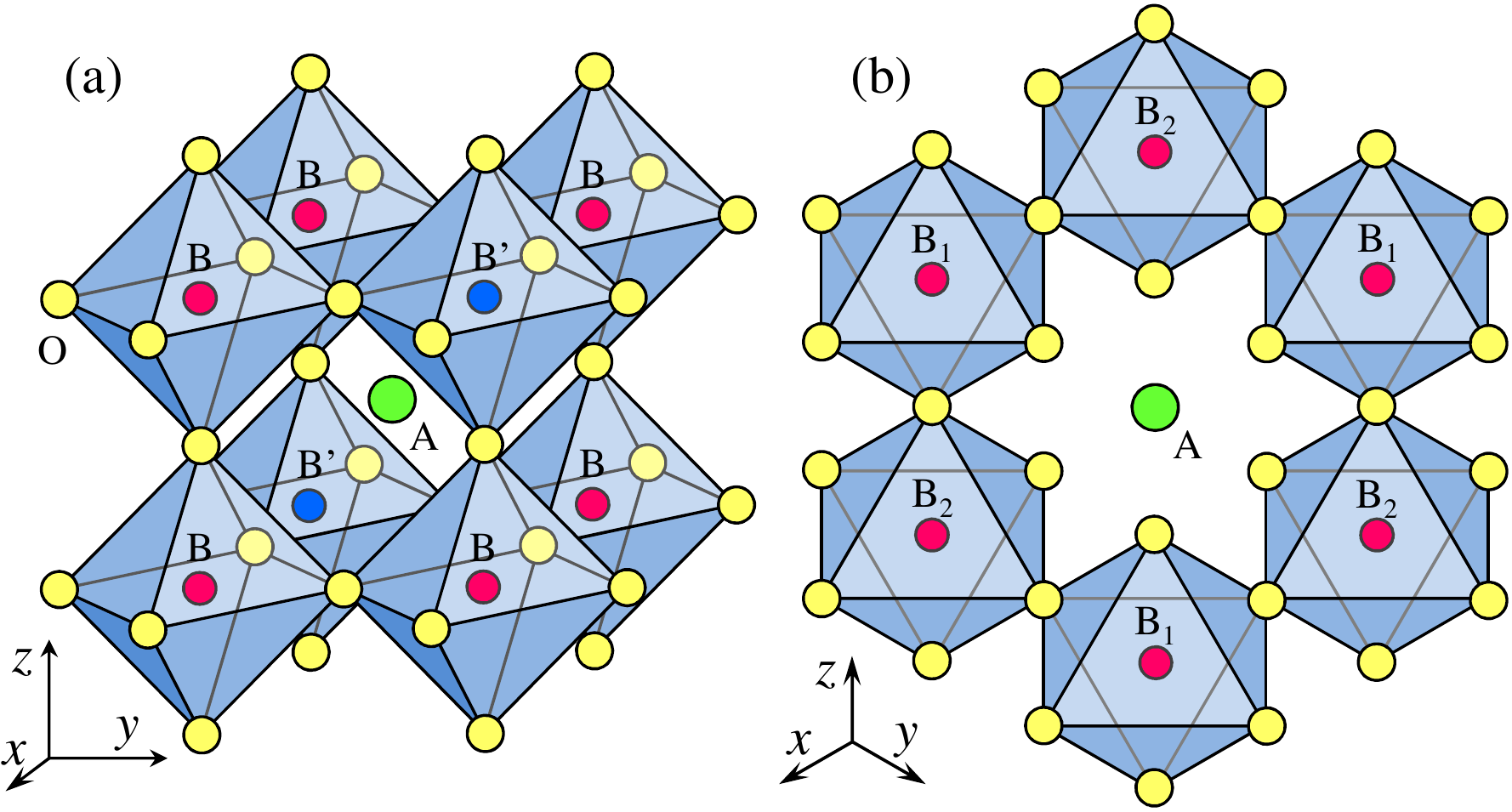}
\caption{(Color online) Buckled honeycomb lattice formed in (111) bilayer of perovskite TMO ABO$_3$. 
(a) (111) bilayer of ABO$_3$ sandwiched by insulating AB'O$_3$. 
(b) ABO$_3$ forms buckled honeycomb lattice with two sublattices B$_1$ and B$_2$. 
}
\label{fig:buckledhoneycomb}
\end{figure}

Recently, based on the tight-binding modeling and subsequent 
density-functional theory (DFT) calculations, 
bilayers of perovskite transition-metal oxides (TMOs) ABO$_3$ grown along the [111] crystallographic axis, i.e., (111) bilayers, 
were proposed as possible candidates for two-dimensional TIs\cite{Xiao11}. 
In such (111) bilayers, buckled honeycomb lattice is formed by B-site transition-metal ions (Fig.~\ref{fig:buckledhoneycomb}). 
TMOs are known to cover a vast extension of unconventional phenomena due to the strong electron-electron and 
electron-lattice interactions \cite{Imada98}.  
Because of the active orbital degrees of freedom, 
the interplay between the SOC and correlation effects could induce phenomena that are absent in abstract theoretical models or those for semiconductors, 
such as a spin-orbit Mott insulator \cite{Kim09}. 
Furthermore, the recent development in synthesizing artificial TMO heterostructures 
provides great tunability over fundamental physical parameters \cite{Hwang12}. 
Once realized in TMOs, 
combining TIs with other novel states would allow us  
to study further novel phenomena utilizing proximity effects involving TIs \cite{Qi08,Fu08}. 
However, while single-electron properties are treated rather accurately using DFT \cite{Xiao11,Lado13}, including the SOC, 
the effect of electron-electron interactions has not been addressed except for limiting cases for such TMO bilayers 
\cite{Xiao11,Ruegg11,Yang11,Ruegg12,Okamoto13,Ruegg13,Doennig13,Doennig14}. 

In this paper, we provide concrete examples of the interplay between the SOC and correlation effects in (111) TMO bilayers. 
As specific examples, we consider SrIrO$_3$ (SIO) and LaAuO$_3$ (LAO), 
for which the previous DFT calculations predicted TI states \cite{Xiao11}, 
and investigate the correlation-induced phase transitions by means of the dynamical-mean-field theory (DMFT) \cite{DMFT}. 
It is shown that LAO bilayer is far from Mott insulating and antiferromagnetic (AF) insulating regimes, 
and a TI state is robust. 
On the contrary, the correlation effect is significant for SIO bilayer, and 
an AF trivial insulating state is realized. 
This is induced by a relatively narrow bandwidth of near-Fermi-level states with the dominant $J_{eff}=1/2$ character, 
which undergoes an orbital-selective topological Mott transition when the magnetic ordering is suppressed. 
Our results indicate that the interplay between the SOC and correlations is diverse, showing the strong material dependence,
and provide further guidelines for studying topological phenomena in TMOs in both bulk and artificial heterostructures. 

The rest of the paper is organized as follows. In Sec. II, %\ref{sec:method}, 
the methodology is described, and in Sec. III %\ref{sec:results} 
numerical results are presented. 
Section IV %\ref{sec:summary} 
is devoted to the summary.

\section{II. Method}
\label{sec:method}

\begin{figure*}[tbp]
\includegraphics[width=1.5\columnwidth,clip]{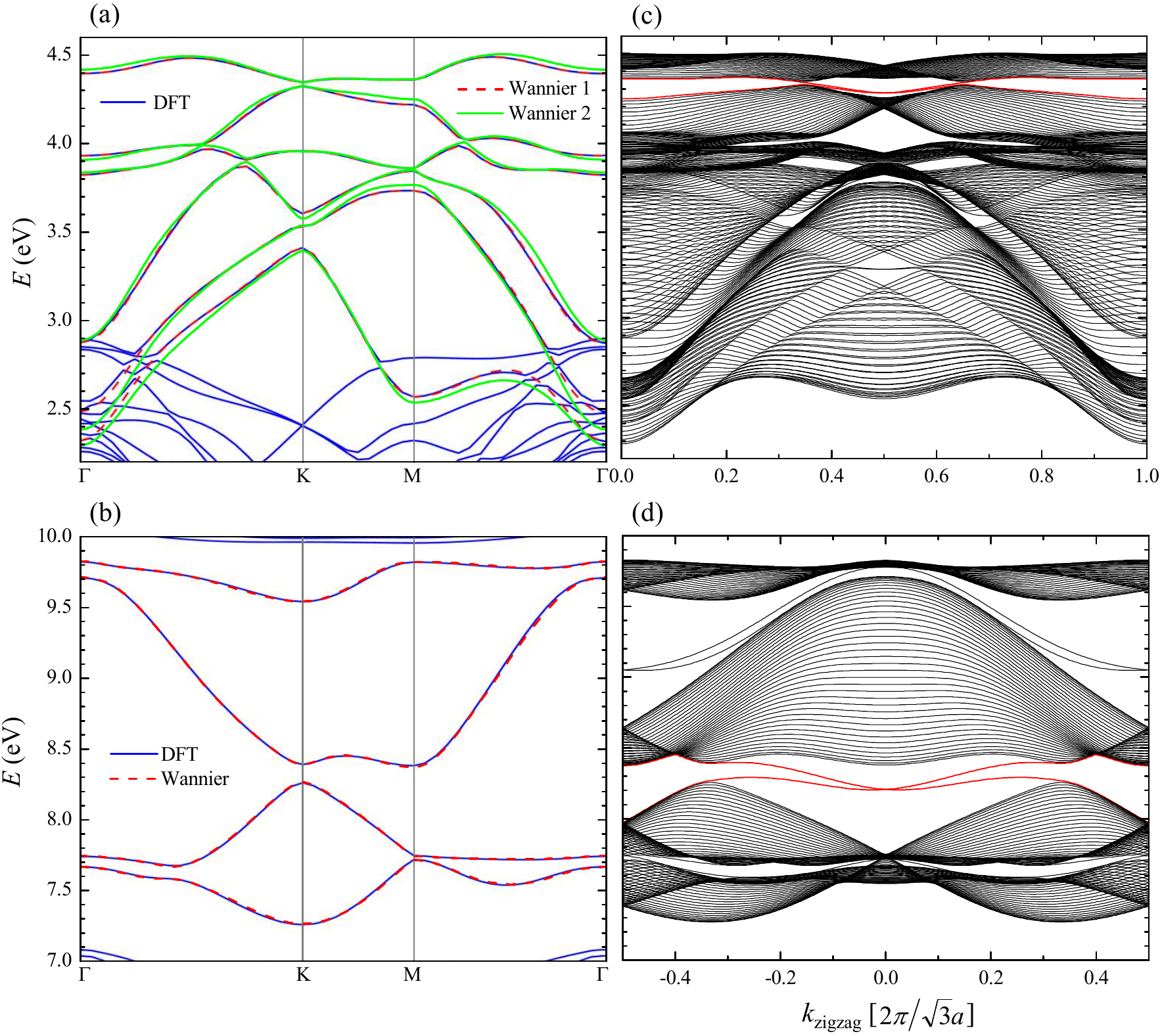}
\caption{(Color online) Bulk dispersion relations (a), (b) and dispersion relations for 80-site thick zigzag slabs (c), (d) of our (111) bilayers. 
(a), (c) SrIrO$_3$ (111) bilayer with the Fermi level located at $E=4.335$ eV. 
(b), (d) LaAuO$_3$ (111) bilayer with the Fermi level at 8.31 eV. 
$a$ is the nearest neighbor bond length projected on the (111) plane. 
}
\label{fig:Wannier}
\end{figure*}

For a realistic treatment on the band effects, 
we generate Wannier functions \cite{Wannier} from the outputs of DFT calculations. 
%using {\it Vienna Ab initio Simulation Package} \cite{VASP} as reported previously.\cite{Xiao11} 
%The details of Wannier functions and comparison with DFT dispersions are presented in Appendix \ref{sec:DFT}. %the supplementary information.\cite{supplementary} 
%
The DFT calculations are performed using {\it Vienna Ab initio Simulation Package} (VASP \cite{VASP}) with 
the use of the projector augmented wave method \cite{Kresse99}
with the generalized gradient approximation in the parametrization of Perdew, Burke and Enzerhof \cite{Perdew96} for exchange correlation. 
%Computational details for DFT calculations are given in Ref.~\onlinecite{Xiao11}. 
In this work, to obtain the Wannier functions, 
thinner supercells are adopted than in Ref.~\cite{Xiao11}, 
consisting of 6 AO$_3$ and 6 B layers along the [111] direction with 
(A,B) = (Sr,Ti) or (La,Al) with two adjacent B layers replaced by TM ions. 
As in the previous study, 
we fix the in-plane lattice parameter to that of bulk SrTiO$_3$ or LaAlO$_3$ and optimize the out-of-plane lattice parameter 
as well as the atomic position 
using 
a $6 \times 6 \times 1$ ${\bf k}$-point mesh including the $\Gamma$ point (0,0,0) for integration over the Brillouin zone.
The SOC is included after the lattice optimization using the default method in VASP. 
We have checked that current DFT results agree with the previous ones.

Figure \ref{fig:Wannier} shows the resultant DFT dispersion relations [(blue) solid lines] and those of Wannier functions. 
With the finite SOC, SrIrO$_3$ (111) bilayer was found to have tiny but finite magnetic moments, 
which prevent us from constructing maximally localized Wannier functions (MLWFs) with proper symmetries, 
whose dispersion relations are given by (red) dashed lines [Fig. \ref{fig:Wannier} (a), Wannier1]. 
To completely suppress the magnetism, we turned off the SOC and derived MLWFs and an effective tight-binding model. 
Then, the following atomic SOC is added to such a model Hamiltonian 
\begin{equation}
H_{SOC} = \frac{\lambda}{2} \sum_{\vec r \sigma \sigma'} \sum_{\tau \tau' \tau''}
i \varepsilon_{\tau \tau' \tau''} d_{\vec r \tau \sigma} \sigma_{\sigma \sigma'}^{\tau''}
d_{\vec r \tau' \sigma'}, 
\end{equation}
where the following convention for the orbital index is used: 
$|a \rangle =|d_{yz} \rangle$, $|b \rangle=| d_{zx} \rangle$, and $|c \rangle=|d_{xy} \rangle$. 
$\sigma^\tau$ with $\tau=a,b,c$ is the Pauli matrix, and 
$\varepsilon_{\tau \tau' \tau''}$ is the Levi-Civita antisymmetric tensor. 
Using $\lambda$ as a fitting parameter, we derive the model Hamiltonian for SrIrO$_3$ (111) bilayer. 
By matching the band gap, we optimized $\lambda$ as 0.36 eV. 
The resulting dispersion relations are given by (green) dash-dot lines [Fig. \ref{fig:Wannier} (a), Wannier2]. 
The data set Wannier2 is used in our realistic DMFT calculation for SrIrO$_3$ (111) bilayer. 
While we did not have issues associated with magnetic moments for LaAuO$_3$ (111) bilayer, 
we found that the symmetry of Wannier functions is progressively reduced by iterations. 
To avoid this, we set the number of iterations to zero, %i.e., {\sf Num\_Iter}=0, 
and derived ``one-shot'' Wannier functions including the SOC. 
The resultant dispersion relations are given by (red) dashed lines [Fig. \ref{fig:Wannier} (b), Wannier]. 
This data set is used in our realistic DMFT calculation for LaAuO$_3$ (111) bilayer.

The dispersion relations for 40-site thick zigzag slab of SrIrO$_3$ (111) bilayer using the Wannier2 parameter set are given in 
Fig. \ref{fig:Wannier} (c). 
Corresponding results for 40-site thick zigzag slab of LaAuO$_3$ (111) bilayer are given in Fig. \ref{fig:Wannier} (d). 
Slab dispersions confirm that both SrIrO$_3$ (111) bilayer and LaAuO$_3$ (111) bilayer are topological insulators 
with gapless edge modes crossing the Fermi level indicated by light (red) lines. 

To examine magnetic ordering, we use structures obtained for paramagnetic phases within DFT as explained above. 
Additionally, we performed DFT+$U$ calculations using the same structures.

Correlation effects are dealt with using DMFT with the exact diagonalization (ED) impurity solver \cite{Caffarel94,Perroni07,ARPACK}. 
Here, the effective medium is approximated as a finite number of bath sites coupled to eigenstates of 
the local non-interacting Hamiltonian \cite{Perroni07}.  
With the SOC and the trigonal crystal field, all degeneracies in $d$ multiplet are lifted except for the Kramers degeneracy \cite{Xiao11}. 
For a $t_{2g}$ system SIO there are three Kramers doublets, 
and for an $e_g$ system LAO there are two Kramers doublets. 
Thus, we introduce two bath sites per Kramers doublet for SIO and three bath sites per doublet for LAO \cite{bath}. 
Note that the highest energy doublet for SIO bilayer has the strong $J_{eff}=1/2$ character, 
while the other two doublets have the $J_{eff}=3/2$ characters. 
As the typical gap amplitude of our systems is $\sim 0.01$-$0.1$ eV, 
we introduce temperature $T=0.01$ eV and 
retain the lowest eigenstates of the interacting impurity Hamiltonians with Boltzmann factors larger than $10^{-5}$. 
The DMFT self-consistency condition is closed by updating the bath Green's functions %${\cal G}^{imp}$ 
at each iteration by minimizing with a conjugate gradient algorithm a distance function that
includes frequency dependence on the discrete Matsubara frequency $\omega_n =  (2n +1) \pi T$
with a frequency weighting $1/|\omega_n|$ \cite{Capone04}. 
The further details of the DMFT including the self-consistency scheme are presented in the Appendix. %the supplementary information.\cite{supplementary} 

\begin{figure}[tbp]
\includegraphics[width=0.8\columnwidth, clip]{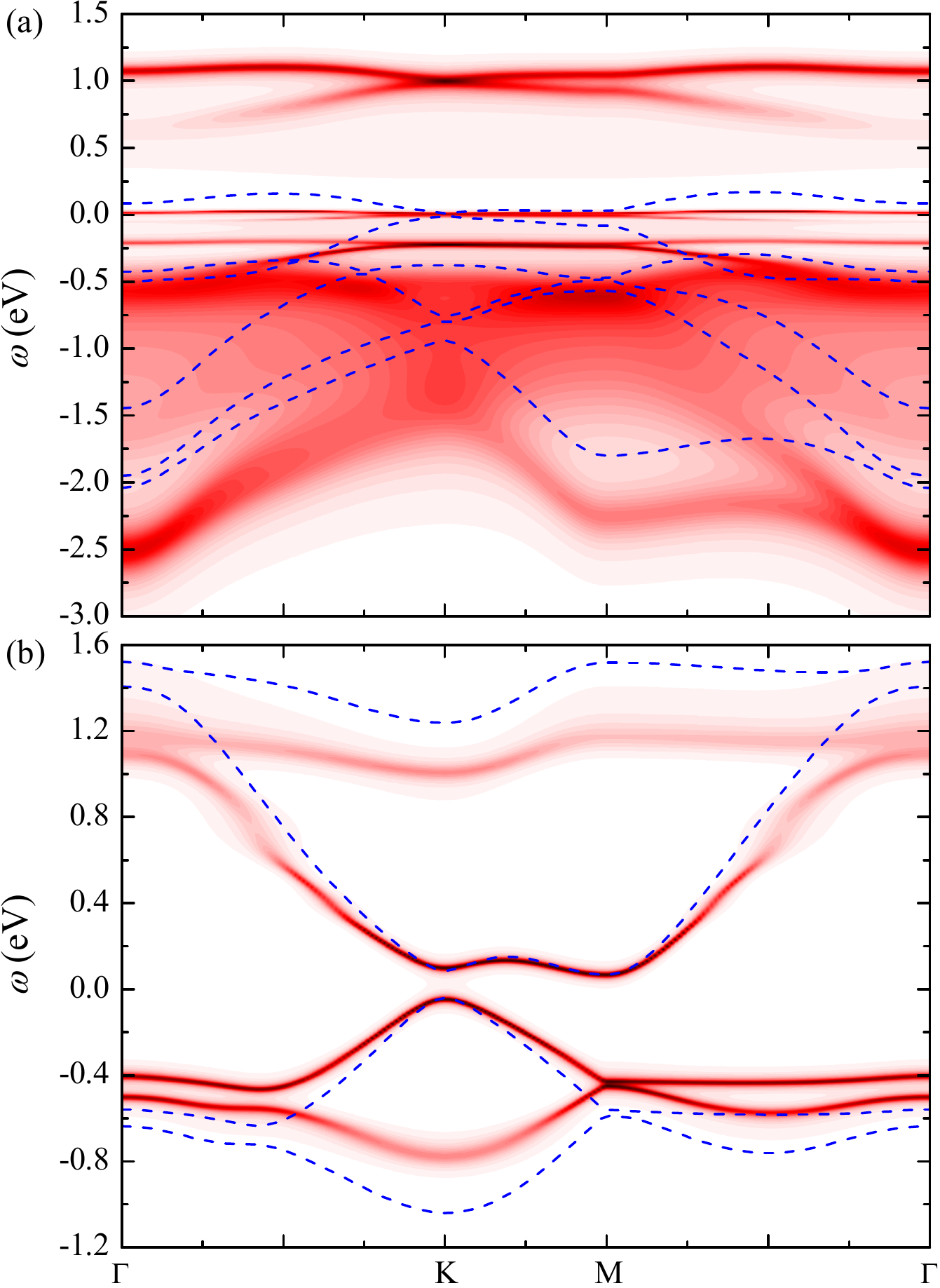}
\caption{(Color online) Bulk spectral function for (a) SrIrO$_3$ bilayer and (b) LaAuO$_3$ bilayer with 
$U=2.0$ and $J=0.2$ eV. 
Dotted lines show dispersion relations of non-interacting models.}
\label{fig:bulkspectra}
\end{figure}

\section{III. Results}
\label{sec:results}

We begin with 
%Figure \ref{fig:bulkspectra} shows 
the typical paramagnetic (PM) bulk spectral functions for (a) SIO bilayer and (b) LAO bilayer 
with $U=2.0$ and $J=0.2$ eV as shown in Fig. \ref{fig:bulkspectra}. 
Here, $U$ is the intraorbital Coulomb interaction and $J$ is the interorbital exchange and pair transfer. 
The interorbital Coulomb interaction is taken to be $U'=U-2J$ throughout the paper. 
The spectral functions are computed using the self-energy directly obtained on the real axis, 
where the $\delta$ function is broadened by using the logarithmic Gaussian function \cite{Sakai89}. 
While both systems have similar bare band width $\sim 2$ eV, the effect of correlations is notably different. 
For SIO bilayer, two nearly-flat dispersions appear at the Fermi level $\omega=0$. 
These bands are mainly coming from the $J_{eff}=1/2$ states, 
but their effective mass is enhanced significantly due to the correlation effects and 
a large amount of spectral weight is transferred to higher energy regimes, 
forming the upper Hubbard band at $\sim 1$ eV and the lower Hubbard band at $\sim -0.3$ eV. 
The other states mainly from $J_{eff}=3/2$ states are relatively unaffected, and 
their dispersion relations are nearly identical to those of non-interacting model with about 0.5 eV downshift 
and broadening due to the imaginary part of the self-energy. 
On the other hand, the spectral function for LAO bilayer is not modified from the non-interacting case 
except for the moderate band renormalization and broadening away from the Fermi level.

\begin{figure}[tbp]
\includegraphics[width=0.8\columnwidth, clip]{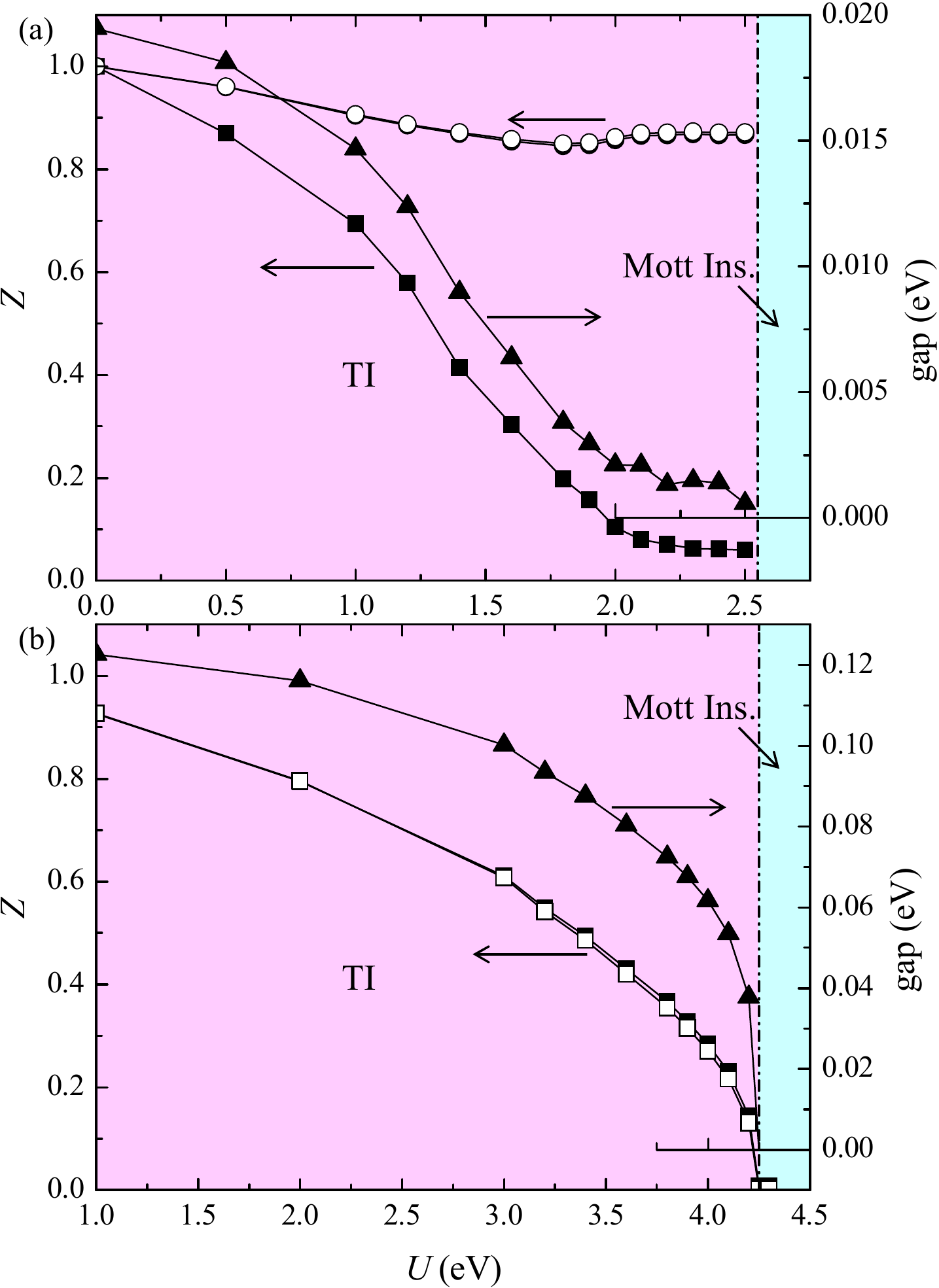}
\caption{(Color online) Quasiparticle weight $Z$ and gap amplitude 
for (a) SrIrO$_3$ bilayer and (b) LaAuO$_3$ bilayer as a function of $U$ with $J/U=0.1$.}
\label{fig:Mott}
\end{figure}

\subsection{A. Mott transition}

In order to see the nature of correlation-induced Mott transition, 
we plot the quasiparticle weight and the gap amplitude of two systems as a function of $U$ in Fig.~\ref{fig:Mott}. 
The quasiparticle weight of Kramers doublet $\alpha$ is evaluated from the self-energy at the lowest Matsubara frequency as 
$Z_\alpha = [1 - \Im \Sigma_\alpha (i \omega_0)/\omega_0]^{-1}$. 
To avoid the broadening of the spectral function by the imaginary part of the self-energy, 
the gap amplitude is evaluated from quasiparticle dispersions \cite{Okamoto11}. 
In both cases, the gap amplitude is monotonically decreased with increasing $U$ and does not show a transition to 
a high $U$ TI state in which the insulating gap increases with $U$ as reported for SIO bilayer in Ref.~\cite{Lado13}. 
On the contrary, the gap amplitude is strongly correlated with the (smallest) quasiparticle weight, 
and they become zero simultaneously at Mott transitions. 
For SIO bilayer, one of three Kramers doublets with the $J_{eff}=1/2$ character undergoes the Mott transition (filled squares) 
while the other two with the $J_{eff}=3/2$ character do not (filled and open circles). 
This situation resembles that in the single-layer perovskite Sr$_2$IrO$_4$ \cite{Arita12}. 
But, the transition in SIO bilayer accompanies the change in the band topology from a nontrivial one to a trivial one 
as confirmed from the fully gapped edge spectra in the latter as discussed later. %shown in Appendix \ref{sec:Mott},%\cite{supplementary} 
Thus, this transition is an orbital-selective topological Mott transition. 
For LAO bilayer, the quasiparticle weights of two Kramers doublets (filled and open squares) show nearly-identical $U$ dependence. 
Whether or not PM Mott insulating states of two systems support gapless ``spinon'' edge states, 
i.e. topological Mott insulators \cite{Pesin10}, 
is a very interesting question but remains beyond the scope of the current single-site DMFT.

%\section{Mott phases} 
%\label{sec:Mott}

\begin{figure}[tbp]
\includegraphics[width=0.8\columnwidth,clip]{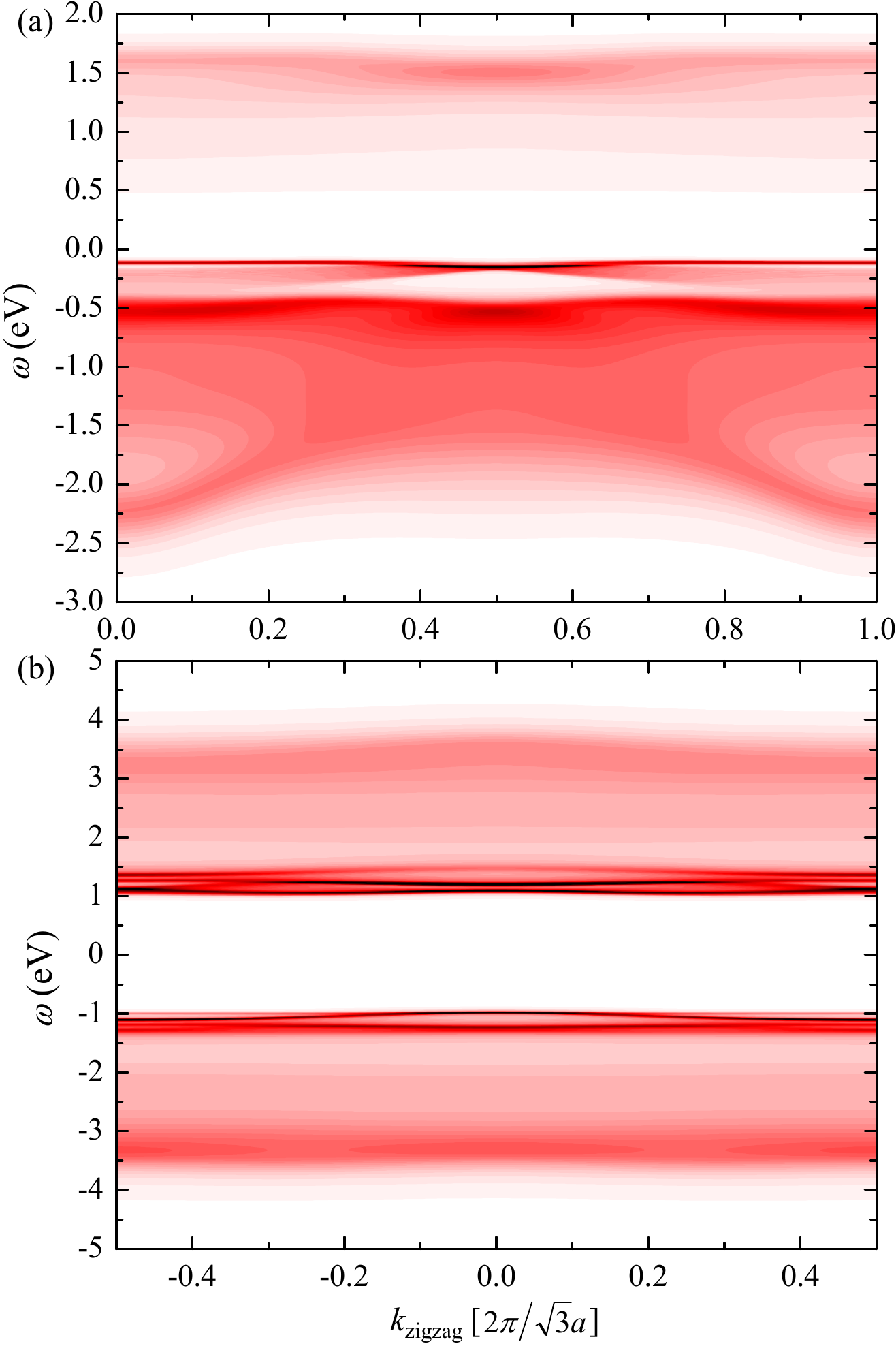}
\caption{(Color online) Edge spectral functions for 40-site thick zigzag slabs of (a) SrIrO$_3$ (111) bilayer with $U=2.6$ eV and $J=0.26$ eV 
and (b) LaAuO$_3$ bilayer with $U=4.4$ eV and $J=0.44$ eV without a magnetic ordering. 
These are in Mott insulating phases. 
}
\label{fig:Mott_slab}
\end{figure}

%Here, we discuss the band topology of our (111) bilayers of TMOs in paramagnetic Mott insulating regimes. 
Figure \ref{fig:Mott_slab} shows the edge spectral functions for 40-site thick zigzag slabs of 
(a) SrIrO$_3$ bilayer with $U=2.6$ eV and $J=0.26$ eV and (b) LaAuO$_3$ bilayer with $U=4.4$ eV and $J=0.44$ eV 
without a magnetic ordering. 
As seen in Fig. \ref{fig:Mott}, 
these are in Mott insulating regimes. 
The absence of edge modes crossing the Fermi level at $\omega=0$ indicates both are in trivial phases at least ``electronically.'' 
There is a possibility that the nontrivial band topology remains in spinon dispersions 
supporting bulk band gaps and gapless edge modes, i.e., topological Mott insulators, 
but the current DMFT does not have access to such states.

\subsection{B. Magnetic ordering}

We next study the effect of magnetic ordering. 
From both a weak-coupling approach and a strong-coupling approach \cite{Okamoto13}, 
a N{\'e}el AF ordering is expected in SIO bilayer. 
Our DFT calculation indeed found the N{\'e}el AF ordering with magnetic moments primarily lying along the [111] axis. 
While the similar ordering is expected in LAO bilayer, 
our DFT calculations with $U$ up to 4 eV and $J/U$ up to $\sim 0.2$ did not find any magnetic ordering. 
This could be ascribed to the fact that the local moment formation is underestimated in DFT. 
Nevertheless, we assume the N{\'e}el AF ordering for both systems by taking the [111] axis, perpendicular to the plane, 
as the spin quantization axis and examine its stability.

\begin{figure}[tbp]
\includegraphics[width=0.8\columnwidth, clip]{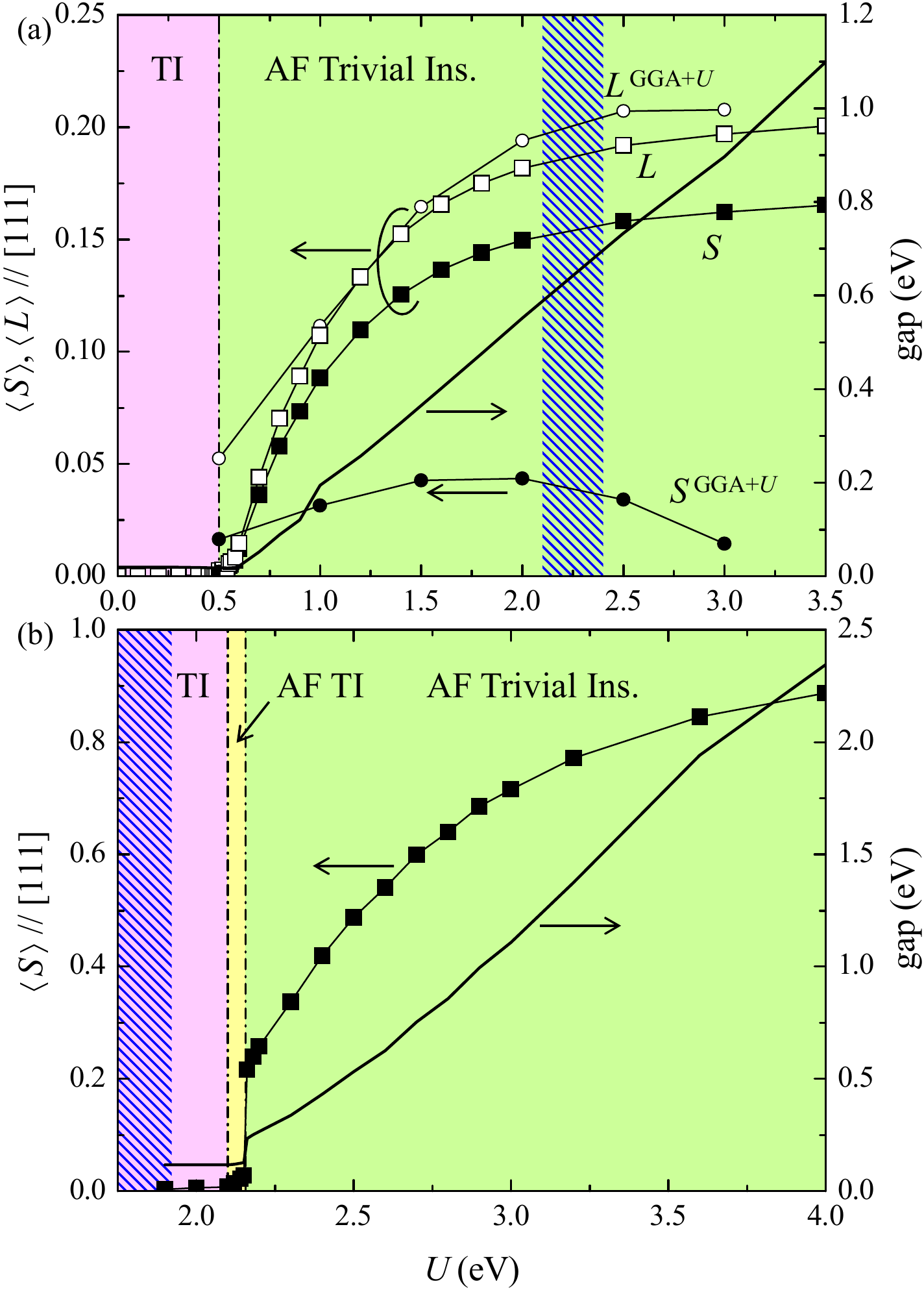}
\caption{(Color online) Staggered spin $(S)$ and angular momentum $(L)$ and gap amplitude 
for (a) SrIrO$_3$ bilayer and (b) LaAuO$_3$ bilayer with antiferromagnetic ordering as a function of $U$ with $J/U=0.1$.
In (a), results of GGA+$U$ are also plotted for spin $S^{{\rm GGA}+U}$ and angular momentum $L^{{\rm GGA}+U}$.
Shaded areas are realistic parameter regimes estimated by cRPA.}
\label{fig:magne}
\end{figure}

Figure \ref{fig:magne} shows the magnetic order parameters and the gap amplitude for (a) SIO and (b) LAO. 
As expected from the smaller critical $U$ for the Mott transition in SIO, 
the critical $U$ for an AF ordering is also small, $U \sim 0.5$ eV. 
In Fig.~\ref{fig:magne} (a), we also plot the magnetic order parameters obtained by generalized gradient approximation (GGA)+$U$. 
The critical $U$ for the AF ordering appears to be slightly smaller than the present DMFT result and 
much smaller than local density approximation +$U$ results in Ref.~\cite{Lado13} where the AF ordering appears at $U>3$ eV. 
In both the DMFT and GGA+$U$ results, 
the orbital moment $\mu_L$ and the spin moment $\mu_S$ are tilted with each other. 
In GGA+$U$, $\mu_S$ is deviated from the [111] direction more strongly than $\mu_L$. 
In DMFT, both $\mu_{L,S}$ turned out to be tilted by a similar amount. 
$|\mu_L|/|\mu_S|$ is found to be $\sim 1.7$ (smaller than 2). 
This indicates the deviation of the spin symmetry from 
the $SU(2)$ point due to the local trigonal field by which the SOC active doublet $e'_g$ is 
lower than the singlet $a_{1g}$. 
On the contrary, $|\mu_L|/|\mu_S|$ becomes larger than 2 at $U > 2$ eV by GGA+$U$. 
%where $\langle S \rangle // [111]$ shows a downturn. 
%This comes from the reduction in $|\mu_S|$ and the increase in the deviation from the [111] axis at large $U$. 
%
In addition, $|\mu_L|$ and $|\mu_S|$ are found to be larger in DMFT than in GGA+$U$. 
These differences are presumably from the better description by DMFT for the local moment formation, 
which becomes important even at small $U$ [see Fig. \ref{fig:Mott} (a)]. 
In contrast to GGA+$U$, 
our DMFT calculation did find the stable AF ordering in LAO at $U \agt 2.1$ eV. 
As a Au$^{+3}$ ion has the $e_g^2$ electron configuration, DFT+$U$ requires very large $J$ to induce local moments $S=1$. 
On the other hand, with finite $J$, $e_g^2$ states with $S=1$ always have larger weights than those with $S=0$. 
These weights increase with increasing $J$, and therefore a magnetic ordering can be induced more easily in DMFT.

In SIO bilayer, an AF N{\'e}el ordering was found to destroy the TI state immediately. 
This is because, in a honeycomb lattice with the AF ordering, 
there is no combined symmetry with the time reversal by which the system remains invariant \cite{Mong10}. 
In LAO bilayer on the other hand, 
there appears a finite window for an AF TI where the gapless edge states and bulk magnetic ordering (time-reversal symmetry breaking) 
coexist as discussed later. 
%(See supplementary material for comparison between the AF TI and the AF trivial insulator.)
%
This comes from the fact that the spin component perpendicular to the [111] plane is conserved in our $e_g$ electron model \cite{Xiao11}, 
and therefore LAO bilayer consists of two copies of Chern insulators with the opposite Chern number. 
%thus the symmetry is higher than that for a $t_{2g}$ electron model for SIO, 
%
The transition between the AF TI phase and the AF trivial insulating phase in LAO bilayer is found to be of the first order 
accompanied by a jump in the staggered magnetic moment. 
Thus, the gap closing is avoided. 
This transition could become a continuous one by including spatial correlations beyond the single-site DMFT. 
For instance, for a correlated BHZ model, 
a single-site DMFT calculation found a discontinuous transition from a PM TI phase to an AF trivial phase \cite{Yoshida13}. 
But a variational cluster approach found a continuous transition accompanied by the gap closing, 
leaving a finite window for an AF TI phase \cite{Miyakoshi13}.

\begin{figure}[tbp]
\includegraphics[width=0.8\columnwidth,clip]{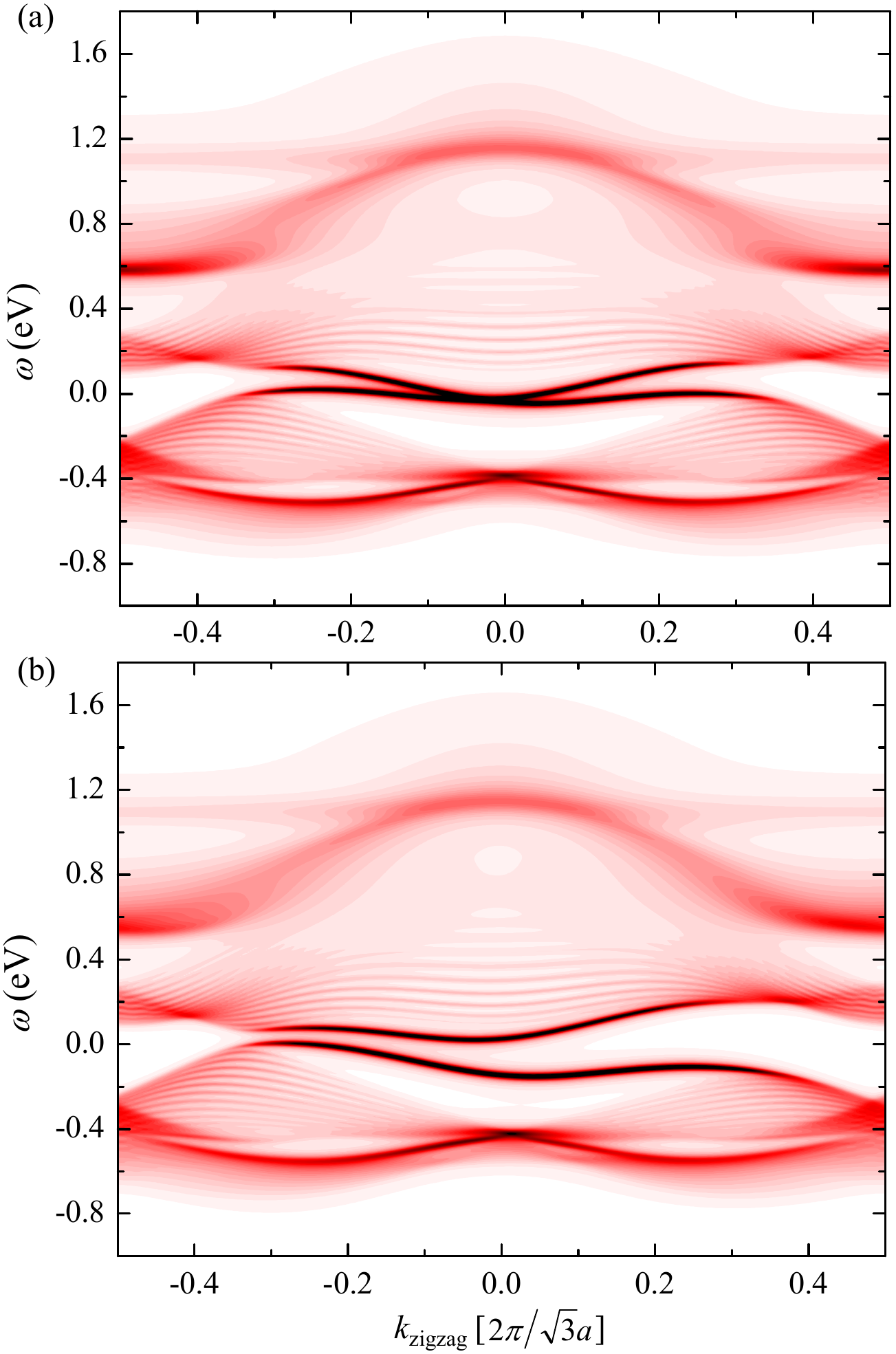}
\caption{(Color online) Edge spectral functions for 40-site thick zigzag slabs of % and (b) 20-site thick armchair slab of 
LaAuO$_3$ (111) bilayer with (a) $U=2.14$ eV %, $J/U=0.1$ and an antiferromagnetic ordering. 
%Corresponding edge spectral functions for $U=2.16$ eV with $J/U=0.1$ are given in (c) and (d). 
and (b) $U=2.16$ eV with $J/U=0.1$. % are given in (c) and (d). 
There are gapless edge modes at the Fermi level $\omega=0$ in (a), while there are none in (b). 
}
\label{fig:AF_slab}
\end{figure}

%\section{Antiferromagnetic phases in $\mbox{LaAuO}_3$ bilayer} 
%\label{sec:AF}
Here, we examine the nature of antiferromagnetic phases in LaAuO$_3$ bilayer in more detail. 
Figure \ref{fig:AF_slab} shows the edge spectral functions for 40-site thick zigzag slabs %and 20-site thick armchair slabs of 
of antiferromagnetic LaAuO$_3$ (111) bilayer with (a) $U=2.14$ eV and (b) $U=2.16$ eV with $J/U=0.1$. 
Because of the N{\'e}el antiferromagnetic ordering, the symmetry between $k>0$ and $k<0$ is broken. % in zigzag slabs. 
While the antiferromagnetic ordering breaks the time-reversal symmetry, 
the combined symmetry between the time reversal and mirror (mirror plane perpendicular to the zigzag direction) remains. 
As a result, each mode is twofold degenerate. 
We see gapless edge modes at the Fermi level in (a) while not in (b), 
indicating the former is in a topologically nontrivial phase but the latter is in a trivial phase. 

\subsection{C. Realistic parameter regimes}

Are SIO and LAO in AF trivial insulating phases or TI phases? 
If they are in AF trivial phases, 
can we turn them into TI phases by suppressing magnetic ordering? 
To answer these questions, we estimate realistic Coulomb interactions by using the constrained random phase approximation
(cRPA)\cite{Aryasetiawan04,Kozhevnikov10,UcRPA,Sakuma13,Taranto13}. 
For SIO bilayer, we took the Slater parameters $F_{0,2,4}$ for Sr$_2$IrO$_4$ from Ref.~\cite{Arita12} and 
deduced $U$ and $J$ 
as 2.232 eV and 0.202 eV, respectively, for the $\{xy, yz, zx\}$ basis. 

\begin{figure}[tbp]
\includegraphics[width=0.7\columnwidth,clip]{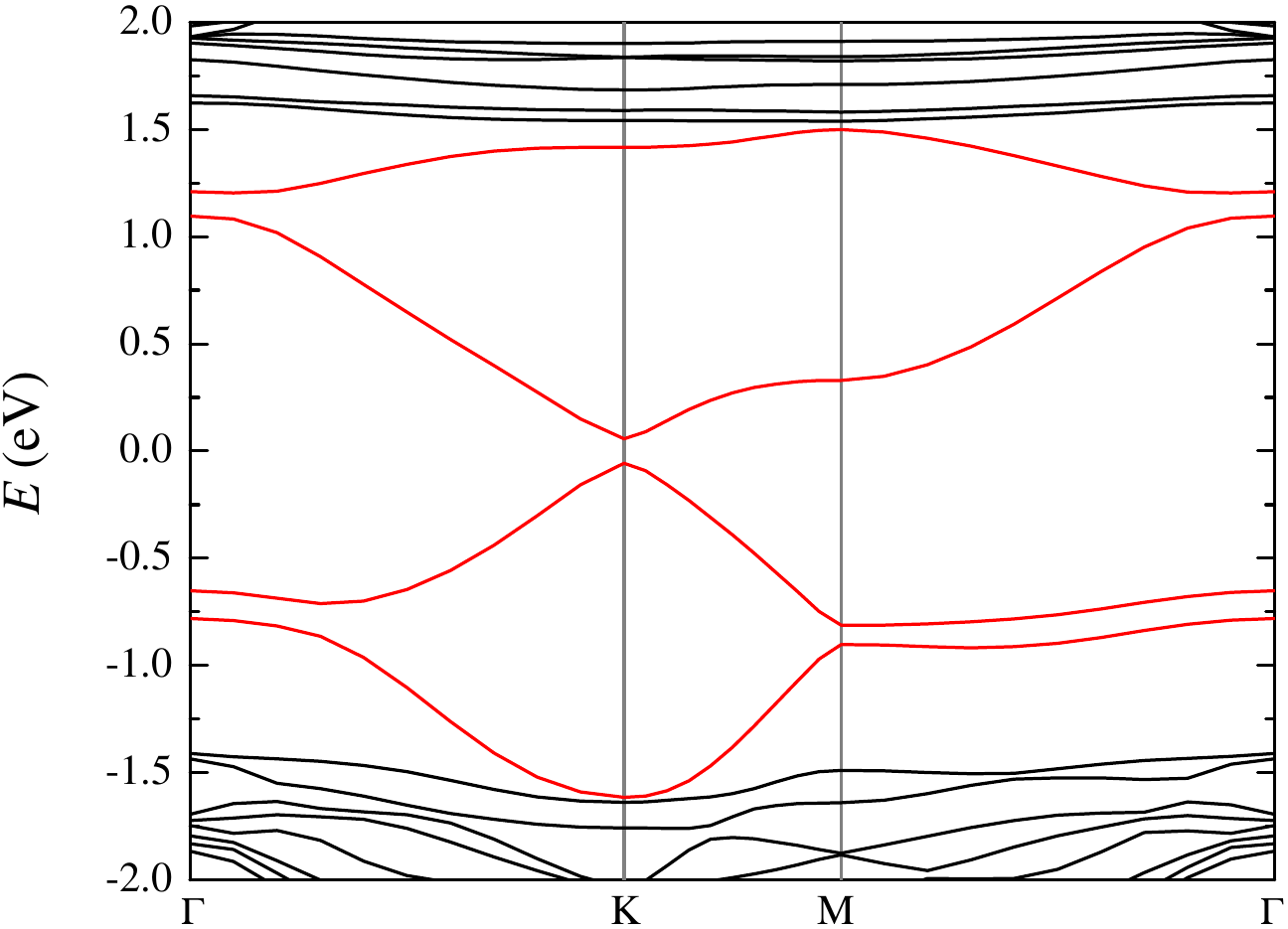}
\caption{(Color online) Bulk dispersion relation for [LaAuO$_3$]$_2$[LaAlO$_3$]$_1$ superlattice, 
which is used to estimate the Coulomb interaction by cRPA.  
Au $5d$ bands are indicated by light (red) lines. 
The Fermi level is located at $E=0$ eV. 
}
\label{fig:LAOcRPA}
\end{figure}

For LAO bilayer, 
we directly computed these parameters for the $\{3z^2-r^2,x^2-y^2\}$ basis.  
%
%Coulomb interactions for LaAuO$_3$ (111) bilayer is estimated by using the cRPA\cite{Aryasetiawan04} 
%as described in Ref.~\onlinecite{Kozhevnikov10}. 
For this purpose, 
we first constructed a thinner supercell consisting of two LaAuO$_3$ and one LaAlO$_3$ layers along the [111] direction in which 
the local structure taken from the previous results with thicker LaAlO$_3$ layers. 
The dispersion relation for the thinner supercell is presented in Fig. \ref{fig:LAOcRPA}. 
We notice that the overall feature is well captured. 
While the Au $5d$ bands are slightly wider than those in Fig. \ref{fig:Wannier} (b), 
these are separated from other bands. 
Thus the realistic $U$ for thicker LAO bilayer might be slightly larger than this supercell, 
but not significantly. 
We then used the density response code for Elk (Ref.~\cite{Kozhevnikov10}). 
%The polarization function was expanded in plane waves with an energy cutoff of 5 Ry 
%and the total number of bands considered in the polarization calculation was set to 80. 
The polarization function was expanded in plane waves with an energy cutoff of 5 Ry and 
the number of empty bands considered in the polarization calculation was set to 80. 
We performed the calculation with $4 \times 4 \times 2$ ${\bf k}$ mesh and $3 \times 3 \times 1$ ${\bf k}$ mesh, 
and we found the difference in the Coulomb interaction parameters is within 5 \%. 
Resultant $U$ and $J$ are $U=1.80$ eV and $J=0.225$ eV, respectively.  
Note that $J/U \approx 0.1$ in both SIO and LAO. 
These realistic parameter regimes are approximately indicated by shades in Fig.~\ref{fig:magne}.

\begin{figure}[tbp]
\includegraphics[width=0.8\columnwidth, clip]{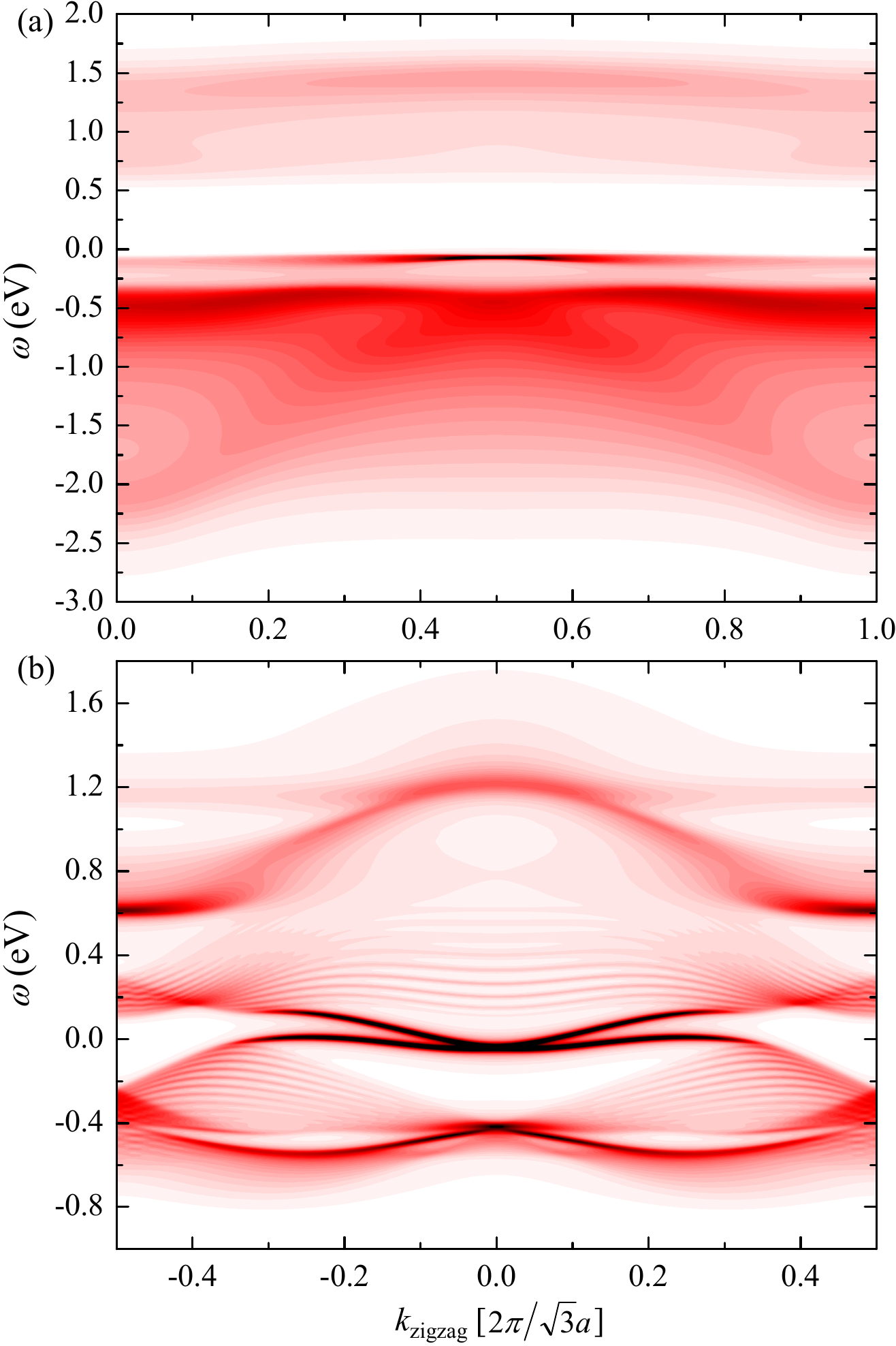}
\caption{(Color online) Edge spectral functions for 40-site thick zigzag slabs of 
(a) SrIrO$_3$ bilayer with $U=2.232$ eV and $J=0.202$ eV and (b) LaAuO$_3$ bilayer with $U=1.80$ and $J=0.225$ eV. 
$a$ is the nearest-neighbor distance projected on the [111] plane. 
Because of an AF ordering, SrIrO$_3$ bilayer does not support gapless edge modes. 
} 
\label{fig:real_slab}
\end{figure}

With the realistic parameters, 
we confirmed that an AF N{\'e}el ordered state is realized in SIO bilayer and a PM state in LAO bilayer. 
From the edge spectra, 
the AF SIO is an AF trivial insulator while the PM LAO is a TI (see Fig.~\ref{fig:real_slab}). 
Note that the critical U for a PM-AF transition is supposed to be underestimated by the current DMFT 
because the long-wave length quantum fluctuation is absent. 
Thus LAO is further away from AF phases. 
When an AF N{\'e}el order is suppressed, SIO bilayer turns into a PM TI not a PM Mott insulator. 
This situation is quite similar to that in Sr$_2$IrO$_4$ \cite{Arita12,Zhang13}. 
Since the gap amplitude of PM SIO bilayer is about 0.002 eV, which is an order of magnitude smaller than temperature $T=0.01$ eV 
[see Fig. \ref{fig:Mott} (a)], 
thermal broadening might hinder the TI nature and SIO bilayer would behave as a topological semimetal. 
In reality, finite spatial correlations may induce a pseudogap even without an AF long-range order in a PM phase of SIO bilayer. 
%with realistic Coulomb interactions. 
Whether or not such a pseudogap phase supports gapless electron or spinon edge modes remains a very interesting problem. 
Even if a pseudogap is created, a high pressure may push SIO bilayer back to a TI. 
%Also, doping carriers into SIO bilayer may induce the unconventional superconductivity.\cite{Okamoto13,Wang11} 
Also, carrier doping may induce the unconventional superconductivity in SIO bilayer \cite{Okamoto13,Wang11}.

\section{IV. Summary}
\label{sec:summary}

To summarize, the electronic properties of (111) TMO bilayers are governed by 
the close interplay between the SOC and correlation effects. 
Using the realistic DMFT, 
the SOC is found to dominate in LAO bilayer, resulting in a moderately correlated TI. 
On the contrary, correlation effects overcome the SOC in SIO bilayer, 
driving it towards an orbital-selective topological Mott transition, 
where only one band with the strong $J_{eff}=1/2$ character %shows the significant quasiparticle renormalization and 
undergoes ``band insulator''--Mott insulator transition. 
SIO bilayer is further unstable against a N{\'e}el AF ordering, resulting in an AF trivial insulator. 
In contrast to a topological Kondo system 
where the nontrivial band topology comes from the hybridization between $f$ states and the conduction band \cite{Wolgast13,Kim13,XZhang13}, 
the correlation effects in TMOs are Mott-Hubbard type 
and $d$ states themselves hold the nontrivial band topology. 
The interplay between the SOC and correlation effects in artificial (111) TMO bilayers is thus diverse 
and brings about rich competing phenomena, 
rendering unique playgrounds for studying such an interplay. 
%Further investigations from both theoretical and experimental side are highly required. 

%\acknowledgements
\section{Acknowledgments}
S.O. thanks M. S. Bahramy, B.-J. Yang and J. Matsuno for their fruitful discussions and 
RIKEN CEMS for hospitality during his visit under the FIRST Theory Forum Visiting Scientist program. 
The research by S.O. is supported by 
the U.S. Department of Energy, Basic Energy Sciences, Materials Sciences and Engineering Division. 
W.Z. is supported by National Natural Science Foundation of China (No. 11374273).
Y.N. is supported by Grant-in-Aid for JSPS Fellows (12J08652). 
D.X. is supported by AFOSR Grant No. FA9550-12-1-0479. 
N.N. is supported by Grant-in-Aids for Scientific Research (S) (No.~24224009) 
from the Ministry of Education, Culture, Sports, Science and Technology (MEXT) of Japan, 
Strategic International Cooperative Program (Joint Research Type) from Japan Science and Technology Agency, 
and by Funding Program for World-Leading Innovative R\&D on Science and Technology (FIRST Program). 

%\begin{widetext}

%\appendix

\section{Appendix: DMFT scheme}
\label{sec:DMFT}

\setcounter{equation}{0}
\renewcommand{\theequation}{A\arabic{equation}}

The DMFT calculations are performed using a single-particle Hamiltonian 
consisting of the local part $H_{local}$ and the nonlocal or band part $H_{nonlocal}$ as parametrized in the Wannier basis and a many-body part $H_U$. 
Here, we express $H_U$ in terms of real orbitals, either $t_{2g}$ or $e_g$, as\cite{Sugano70} 
\begin{eqnarray}
H_U \!\! &=& \!\! U\sum_\alpha d^\dag_{\alpha \uparrow} d_{\alpha \uparrow} d^\dag_{\alpha \downarrow} d_{\alpha \downarrow}
+U' \sum_{\alpha \ne \beta} d^\dag_{\alpha \uparrow} d_{\alpha \uparrow} d^\dag_{\beta \downarrow} d_{\beta \downarrow} \nonumber \\
&&\!\! +(U'-J) \!\! \sum_{\alpha > \beta, \sigma} \!\! d^\dag_{\alpha \sigma} d_{\alpha \sigma} d^\dag_{\beta \sigma} d_{\beta \sigma} \nonumber \\
&&\!\! + J \sum_{\alpha \ne \beta} \Bigl( 
d^\dag_{\alpha \uparrow} d_{\beta \uparrow} d^\dag_{\beta \downarrow}d_{\alpha \downarrow}
+ d^\dag_{\alpha \uparrow} d_{\beta \uparrow} d^\dag_{\alpha \downarrow} d_{\beta \downarrow} \Bigr), 
\end{eqnarray}
where $\alpha$ and $\beta$ stand for $(yz,zx,xy)$ for SIO and $(3z^2-r^2,x^2-y^2)$ for LAO.  
$U$ and $U'$ are the intraorbital Coulomb interaction and the interorbital Coulomb interaction, respectively, 
and $J$ represents the interorbital exchange interaction (fourth term) and the interorbital pair hopping (fifth term). 
For orbitals with the $t_{2g}$ or $e_g$ symmetry, $U'=U-2J$. 
We then move on to the basis which diagonalizes the local non-interacting Hamiltonian $H_{local}$ including the crystal field and the SOC, 
say $d_{\xi \tau} = \sum_{\alpha \sigma} W^*_{\alpha \sigma, \xi\tau} d_{\alpha \sigma}$ with the eigenvalue $\varepsilon_{\xi \tau}$. 
$\tau=\pm$ specifies a pair of Kramers doublet $\xi$. 
In the current ED-DMFT, the effective medium is approximated as a finite number of bath sites coupled to each Kramers doublet. 
Thus, the impurity model to be diagonalized is written as 
\begin{eqnarray}
\hspace{-1.5em} H_{imp} \!\!&=&\!\! H_{local} + H_U 
+  \sum_{\xi \tau} \sum_{l=1}^{n_s} \nonumber \\
&&\!\! \times \Bigl[ \varepsilon_{\xi \tau l} c^\dag_{\xi \tau l} c_{\xi \tau l} 
 + \bigl( V_{\xi \tau l} c^\dag_{\xi \tau l} d_{\xi \tau} + h.c. \bigr) \Bigr], 
% &=& \!\! H_{local} + H_U 
%+ \sum_{\xi \tau} \sum_{l=1}^{n_s} \Bigl\{ \varepsilon_{\xi \tau l} c^\dag_{\xi \tau l} c_{\xi \tau l}  \nonumber \\
%&&+ \sum_{\alpha \sigma} \bigl( V_{\xi \tau l} c^\dag_{\xi \tau l} W_{\alpha \sigma, \xi\tau} d_{\alpha \sigma} + h.c. \bigr) \Bigr\} , 
\end{eqnarray}
where $n_s$ is taken to be 2 for the $t_{2g}$ system SIO and 3 for the $e_g$ system LAO.

By solving $H_{imp}$, one obtains the interacting Green's function $G_{\xi \tau} (i \omega_n)$ and 
the self-energy $\Sigma_{\xi \tau} (i \omega_n) = {\cal G}^{-1}_{\xi \tau} (i \omega_n) - G^{-1}_{\xi \tau} (i \omega_n)$, 
where the non-interacting Green's function is given by 
\begin{equation}
{\cal G}^{-1}_{\xi \tau} (i \omega_n) = i \omega  - \varepsilon_{\xi \tau} - \sum_l \frac{|V_{\xi \tau l}|^2}{i \omega_n - \varepsilon_{\xi \tau l}}. 
\end{equation}
Parameters $\varepsilon_{\xi \tau l}$ and $V_{\xi \tau l}$ are fixed at each DMFT iteration 
by minimizing with a conjugate gradient algorithm a distance function 
\begin{equation}
\Delta = \sum_{\xi \tau} \sum_n \Bigl| {\cal G}_{\xi \tau} (i \omega) - g^0_{\xi \tau} (i \omega_n)  \Bigr|^2\frac{1}{|\omega_n|}. 
\end{equation}
Here,  $g^0_{\xi \tau} (i \omega_n) $ is the lattice non-interacting Green's function, 
which is obtained by 
\begin{equation}
g^0_{\xi \tau} (i \omega_n) = \Bigl[ g^{-1}_{\xi \tau} (i \omega_n)  + \Sigma_{\xi \tau} (i \omega_n) \Bigr]^{-1},  
\end{equation}
with the lattice interacting Green's function 
\begin{equation}
g_{\xi \tau} (i \omega_n) = \sum_{\alpha \beta \sigma \sigma'} W^*_{\alpha \sigma, \xi \tau} g_{\alpha \sigma, \beta \sigma'} (i \omega_n) W_{\beta \sigma', \xi \tau}.   
\end{equation}
Now, reviving the sublattice index $\iota = 1,2$, $g^{(\iota)}_{\alpha \sigma, \beta \sigma'} (i \omega_n) $ is formally given by 
\begin{widetext}
\begin{eqnarray}
g^{(\iota)}_{\alpha \sigma, \beta \sigma'} (i \omega_n) 
=
\int 
\biggl( \frac{dk}{2\pi} \biggr)^2 \!\! \left[\frac{1}{ i \omega_n - \hat H_{local} - \hat H_{nonlocal} (\vec k) - \hat \Sigma (i \omega_n)} \right]_{\iota \alpha \sigma, \iota \beta \sigma'}. 
\end{eqnarray}
\end{widetext}
Here, the lattice self-energy matrix is expressed as 
\begin{eqnarray}
\hat \Sigma (i \omega_n) = 
\left[ \begin{array}{cc}
\hat \Sigma^{(1)} (i \omega_n) &   \\
  & \hat \Sigma^{(2)}  (i \omega_n)
\end{array} \right] , 
\end{eqnarray}
with 
$\hat \Sigma^{(\iota)} (i \omega_n)$ given by 
\begin{equation}
\Sigma^{(\iota)}_{\alpha \sigma, \beta \sigma'} (i \omega_n) = W_{\alpha \sigma, \xi \tau} \Sigma^{(\iota)}_{\xi \tau} W^*_{\beta \sigma', \xi \tau}. 
\end{equation}
By symmetry, $\Sigma^{(\iota)}_{\xi \tau} (i \omega_n) = \Sigma_{\xi} (i \omega_n)$ in a PM phase and 
$\Sigma^{(1)}_{\xi \tau} (i \omega_n) = \Sigma^{(2)}_{\xi -\tau} (i \omega_n)$ in an AF phase.


\begin{thebibliography}{*}
\bibitem{UcRPA}
For $3d$ TMOs, LDA+DMFT with Coulomb interactions by cRPA tends to underestimate the mass enhancement \cite{Sakuma13} 
because the LDA bare band is wide. 
This issue is resolved in $GW$+DMFT due to the additional band renormalization introduced by $GW$ \cite{Taranto13}. 
As our systems are $5d$ TMOs, the Wannier functions are more extended and 
such effects are expected to be small.
\end{thebibliography}
\end{document}